%% file: main.tex
\documentclass{article}
\usepackage{spconf}

\input{miko}
\pgfplotsset{compat=1.15}

\Crefname{figure}{Fig.}{Figs.}

\newcommand{\hhat}{\hat{\mbh}}

\newacronym{amp}{AMP}{approximate message passing}
\newacronym{cnn}{CNN}{convolutional neural network}
\newacronym{dft}{DFT}{discrete Fourier transform}
\newacronym{em}{EM}{expectation-maximization}
\newacronym{gmm}{GMM}{Gaussian mixture model}
\newacronym{lmmse}{LMMSE}{linear minimum mean square error}
\newacronym{los}{LOS}{line of sight}
\newacronym{ls}{LS}{least squares}
\newacronym{mimo}{MIMO}{multiple-input multiple-output}
\newacronym{mse}{MSE}{mean square error}
\newacronym{nlos}{NLOS}{non-line of sight}
\newacronym{o2i}{O2I}{outdoor-to-indoor}
\newacronym{omp}{OMP}{orthogonal matching pursuit}
\newacronym{pdf}{PDF}{probability density function}
\newacronym{simo}{SIMO}{single-input multiple-output}
\newacronym{siso}{SISO}{single-input single-output}
\newacronym{snr}{SNR}{signal-to-noise ratio}
\newacronym{ula}{ULA}{uniform linear array}

\newtheorem{theorem}{Theorem}

\newcommand{\legendAmp}{\footnotesize AMP}
\newcommand{\legendCnn}{\footnotesize CNN}
\newcommand{\legendGenielmmse}{\footnotesize gen. LMMSE}
\newcommand{\legendGlobalcov}{\footnotesize sample cov.}
\newcommand{\legendGmm}{\footnotesize GMM}
\newcommand{\legendLs}{\footnotesize LS}
\newcommand{\legendOmp}{\footnotesize gen. OMP}

\newcommand{\plotheightcomponents}{0.6\columnwidth}
\newcommand{\plotheightgpp}{0.68\columnwidth}
\newcommand{\plotheightquadriga}{0.58\columnwidth}
\newcommand{\plotwidth}{0.95\columnwidth}

\newcommand{\lineWidth}{1.0pt}
\newcommand{\marksize}{1.8pt}
\tikzset{amp/.style={mark options={solid},color=TUMBeamerYellow, line width=\lineWidth, mark size=\marksize, dashdotted}}
\tikzset{cnn/.style={mark options={solid},color=black, line width=\lineWidth, mark=x, mark size=\marksize, dotted}}
\tikzset{genielmmse/.style={mark options={solid},color=TUMBeamerRed, line width=\lineWidth, mark=diamond, mark size=\marksize, dashed}}
\tikzset{globalcov/.style={mark options={solid},color=TUMBeamerOrange, line width=\lineWidth, mark=o, mark size=\marksize, dashed}}
\tikzset{gmm/.style={mark options={solid},color=TUMBeamerBlue, line width=\lineWidth, mark=square, mark size=\marksize, dashed}}
\tikzset{ls/.style={mark options={solid},color=TUMMediumGray, line width=\lineWidth, mark=pentagon, mark size=\marksize, dashed}}
\tikzset{omp/.style={mark options={solid},color=TUMBeamerGreen, line width=\lineWidth, mark=triangle, mark size=\marksize, dashed}}

\tikzset{onePlow/.style={mark options={solid},color=TUMBeamerBlue, line width=\lineWidth, mark=square, mark size=\marksize, solid}}
\tikzset{onePhigh/.style={mark options={solid},color=TUMBeamerBlue, line width=\lineWidth, mark=square, mark size=\marksize, dashed}}
\tikzset{threePlow/.style={mark options={solid},color=black, line width=\lineWidth, mark=x, mark size=\marksize, solid}}
\tikzset{threePhigh/.style={mark options={solid},color=black, line width=\lineWidth, mark=x, mark size=\marksize, dashed}}
\tikzset{quaDlow/.style={mark options={solid},color=TUMBeamerRed, line width=\lineWidth, mark=diamond, mark size=\marksize, solid}}
\tikzset{quaDhigh/.style={mark options={solid},color=TUMBeamerRed, line width=\lineWidth, mark=diamond, mark size=\marksize, dashed}}

\title{An asymptotically optimal approximation of the conditional mean channel estimator based on Gaussian mixture models}

\name{Michael Koller, Benedikt Fesl, Nurettin Turan, Wolfgang Utschick}
\address{Technische Universit\"at M\"unchen}
\begin{document}
	
\glsdisablehyper
\maketitle
\begin{abstract}
This paper investigates a channel estimator based on \acp{gmm}.
We fit a \ac{gmm} to given channel samples to obtain an analytic \ac{pdf} which approximates the true channel \ac{pdf}.
Then, a conditional mean channel estimator corresponding to this approximating \ac{pdf} is computed in closed form and used as an approximation of the optimal conditional mean estimator based on the true channel \ac{pdf}.
This optimal estimator cannot be calculated analytically because the true channel \ac{pdf} is generally not available.
To motivate the \ac{gmm}-based estimator, we show that it converges to the optimal conditional mean estimator as the number of \ac{gmm} components is increased.
In numerical experiments, a reasonable number of \ac{gmm} components already shows promising estimation results.
\end{abstract}
\begin{keywords}
conditional mean channel estimation,
Gaussian mixture models,
expectation-maximization,
machine learning,
spatial channel model
\end{keywords}
%

\glsresetall

\section{INTRODUCTION}
\begin{figure}[b]
	\onecolumn
	\centering
	\copyright This work has been submitted to the IEEE for possible publication. Copyright may be transferred without notice, after which this version may no longer be accessible.
	
	\vspace{-2.05cm}
	\twocolumn
\end{figure}

Channel estimation plays an important role in future mobile communications systems, e.g., \cite{RuPeLaLaMaEdTu13,ArDeAxMo14,SaRa16}.
For many system models, it is known that a conditional mean channel estimator yields \ac{mse} minimizing estimates.
However, computing the conditional mean estimator in closed form requires analytic knowledge of the channel \ac{pdf}, which is generally not available.
Even if the \ac{pdf} is given, calculating the conditional mean estimator might not be possible analytically or not be tractable practically.
Increasingly, advanced channel models (e.g., \cite{3gpp}) or simulators (e.g., \cite{QuaDRiGa1, QuaDRiGa2}) are used to generate large amounts of realistic channel data and the goal is to design estimation algorithms based on these.
Data sets are also available for download~\cite{Al19} or can be obtained via a measurement campaign.

The main idea of this paper is to use the available data to approximate the unknown channel \ac{pdf} by means of a \ac{gmm}.
This is motivated by the good approximation properties of \acp{gmm}~\cite{NgNgChMc20} and the increasing need to analyze collected data from measurement campaigns and available channel simulation tools or to characterize them by means of providing optimal estimators.
Given a \ac{gmm} approximation of the true channel \ac{pdf}, we can analytically compute a corresponding conditional mean estimator, which is then used for the actual channel estimation.
We show that this \ac{gmm}-based estimator converges to the optimal conditional mean estimator as the number of \ac{gmm} components increases.

In \cite{WeJiWoChTi15,LiZhTaTi19,GuZh19}, e.g., \acp{gmm} are used to approximate or model the true channel \ac{pdf}, e.g., for pilot design or channel clustering.
To the best of our knowledge, the obtained \acp{gmm} have not been used to approximate the conditional mean estimator asymptotically optimally, which is studied in this work.


\section{SIGNAL MODEL AND OPTIMAL ESTIMATOR}

We consider a single-antenna user who transmits pilots to an \( N \)-antennas base station which receives
\begin{equation}\label{eq:signal_model}
    \C^N \ni \mby = \mbh + \mbn
\end{equation}
where \( \mbh \in \C^N \) is the channel and \( \mbn \) is additive white Gaussian noise with mean zero and covariance \( \mbSigma \in \C^{N\times N} \).
We denote the noise \ac{pdf} by \( f_{\mbn} \).
The channel is assumed to be described by means of a continuous \ac{pdf} \( f_{\mbh} \).

It is well known that the \ac{mse}-optimal channel estimator is given by the conditional mean estimator
\begin{equation}\label{eq:conditional_mean}
    \hhat = \expec[\mbh \mid \mby] = \int \mbh f_{\mbh\mid\mby}(\mbh\mid\mby) d \mbh.
\end{equation}
The conditional \ac{pdf} of the channel given the observation is
\begin{equation}\label{eq:fh_given_y}
    f_{\mbh\mid\mby}(\mbh\mid\mby) = \frac{f_{\mby\mid\mbh}(\mby\mid\mbh)f_{\mbh}(\mbh)}{f_{\mby}(\mby)} = \frac{f_{\mbn}(\mby-\mbh)f_{\mbh}(\mbh)}{f_{\mby}(\mby)}
\end{equation}
where \( f_{\mby} \) is the \ac{pdf} of the receive signal.
This paper proposes a method to approximate~\eqref{eq:conditional_mean} with the help of \acp{gmm}.

\section{CHANNEL ESTIMATION WITH GAUSSIAN MIXTURE MODELS}\label{sec:estimator}

The optimal estimator~\eqref{eq:conditional_mean} can generally not be computed analytically, but we can assume to have access to a set \( \calH_M = \{ \mbh_m \}_{m=1}^M \) of channel samples.
The following summarizes the proposed procedure to obtain a channel estimator using \( \calH_M \) and the details can be found in~\Cref{sec:channel_estimator}.

Motivated by universal approximation properties of \acp{gmm} \cite{NgNgChMc20}, we fit a \ac{gmm} \( f_{\mbh}^{(K)} \) with \( K \) components to the given data \( \calH_M \) to approximate the unknown channel \ac{pdf} \( f_{\mbh} \).
Then, we define a conditional mean estimator \( \hhat^{(K)} \) as in~\eqref{eq:conditional_mean} but we replace \( f_{\mbh\mid\mby} \) with \( f_{\mbh\mid\mby}^{(K)} \) which is the conditional density based on the approximation \( f_{\mbh}^{(K)} \) instead of the true \ac{pdf} \( f_{\mbh} \).
By universal approximation, \( f_{\mbh}^{(K)} \) converges to the true channel \ac{pdf} \( f_{\mbh} \) for \( K \to \infty \).
We show that this implies the convergence of the proposed estimator \( \hhat^{(K)} \) to the optimal estimator \( \hhat \) for \( K \to \infty \).
Lastly, we present how \( \hhat^{(K)} \), in contrast to \( \hhat \), can always be expressed in closed form.

\subsection{Channel Estimator}\label{sec:channel_estimator}

By \cite[Theorem 5]{NgNgChMc20} (universal approximation property of \acp{gmm}), there exists a sequence \( (f_{\mbh}^{(K)})_{K=1}^\infty \) of \acp{gmm} which converges to the channel \ac{pdf} \( f_{\mbh} \) in the sense that
\begin{equation}\label{eq:convergence_fh}
    \lim_{K\to\infty} \| f_{\mbh} - f_{\mbh}^{(K)} \|_\infty = 0
\end{equation}
holds.
For every \( K \), we define an estimator
\begin{equation}\label{eq:conditional_mean_K}
    \hhat^{(K)} = \expec^{(K)}[\mbh \mid \mby] = \int \mbh f_{\mbh\mid\mby}^{(K)}(\mbh\mid\mby) d \mbh
\end{equation}
where, in analogy to~\eqref{eq:fh_given_y}, we have
\begin{equation}\label{eq:fh_given_y_K}
    f_{\mbh\mid\mby}^{(K)}(\mbh\mid\mby) = \frac{f_{\mbn}(\mby-\mbh)f_{\mbh}^{(K)}(\mbh)}{f_{\mby}^{(K)}(\mby)}.
\end{equation}
Here, \( f_{\mby}^{(K)} \) is the \ac{pdf} of the receive signal~\eqref{eq:signal_model} in the case where the channel \( \mbh \) is distributed according to \( f_{\mbh}^{(K)} \).
With this notation, we have the following theorem.
\begin{theorem}\label{thm:main_result}
    For any \( \mby \in \C^N \), the estimator~\eqref{eq:conditional_mean_K} converges to the optimal estimator~\eqref{eq:conditional_mean} in the sense that it holds:
    \begin{equation}\label{eq:convergence_estimator}
        \lim_{K\to\infty} \| \hhat - \hhat^{(K)} \| = 0.
    \end{equation}
\end{theorem}
This theorem gives a strong motivation to use~\eqref{eq:conditional_mean_K} as a channel estimator for some fixed \( K \).
Due to space limitations, we only sketch the proof of the theorem in~\Cref{sec:sketch} and provide the details in a future work.
Before we turn to the sketch, we explain how to compute~\eqref{eq:conditional_mean_K} in closed form in practice.

A \ac{gmm} with \( K \) components is a \ac{pdf} of the form
\begin{equation}\label{eq:gmm_h}
    f_{\mbh}^{(K)}(\mbh) = \sum_{k=1}^K p(k) \calN_{\C}(\mbh; \mbmu_k, \mbC_k).
\end{equation}
with \( K \) \textit{mixing coefficients} \( p(k) \) as well as \( K \) means \( \mbmu_k \in \C^N \) and covariances \( \mbC_k \in \C^{N\times N} \).
Here, \( \calN_{\C}(\mbh; \mbmu_k, \mbC_k) \) is a Gaussian \ac{pdf} with mean \( \mbmu_k \) and covariance \( \mbC_k \) evaluated at \( \mbh \).
An \ac{em} algorithm in conjunction with given data samples can be used to obtain the parameters in~\eqref{eq:gmm_h}.
An introduction to \acp{gmm} can, e.g., be found in~\cite{bookBi06}.

With a \ac{gmm}~\eqref{eq:gmm_h} given, it remains to calculate the estimator \( \hhat^{(K)} \) in~\eqref{eq:conditional_mean_K}.
To this end, we write~\eqref{eq:conditional_mean_K} as~\cite{FlChKaEk12}
\begin{equation}\label{eq:total_expectation}
    \hhat^{(K)} = \sum_{k=1}^K p(k\mid \mby) \expec^{(K)}[\mbh \mid \mby, k].
\end{equation}
In this expression, \( \mbh \) is distributed according to \( f_{\mbh}^{(K)} \).
This implies that conditioned on one of the components \( k \), the observation \( \mby \mid k \) is a sum of two Gaussian random variables (\( \mbh \mid k \) plus noise \( \mbn \)).
Hence, the conditional expectation \( \expec^{(K)}[\mbh \mid \mby, k] \) can be computed using the well-known \ac{lmmse} formula
\begin{equation}\label{eq:lmmse_formula}
    \expec^{(K)}[\mbh \mid \mby, k] = \mbC_k (\mbC_k + \mbSigma)^{-1} (\mby - \mbmu_k) + \mbmu_k.
\end{equation}
Additionally, we obtain the \ac{pdf} of \( \mby \) as
\begin{equation}
    f_{\mby}^{(K)}(\mby) = \sum_{k=1}^K p(k) \calN_{\C}(\mby; \mbmu_k, \mbC_k + \mbSigma).
\end{equation}
This is a \ac{gmm} as well.
\acp{gmm} generally allow to compute the so-called \textit{responsibilities} by evaluating Gaussian \acp{pdf}~\cite{bookBi06}:
\begin{equation}\label{eq:responsibilities}
    p(k \mid \mby) = \frac{p(k) \calN_{\C}(\mby; \mbmu_k, \mbC_k + \mbSigma)}{\sum_{i=1}^K p(i) \calN_{\C}(\mby; \mbmu_i, \mbC_i + \mbSigma) }.
\end{equation}
Summarily, plugging~\eqref{eq:responsibilities} and~\eqref{eq:lmmse_formula} into~\eqref{eq:total_expectation} yields a closed-form expression for the channel estimator \( \hhat^{(K)} \).

\subsection{Proof Sketch}\label{sec:sketch}

Note that because \acp{gmm} can approximate any continuous \ac{pdf} arbitrarily well~\cite{NgNgChMc20}, there always exists a sequence of \acp{gmm} which converges to \( f_{\mbh\mid\mby} \) for any given observation \( \mby \).
In light of this, proving~\Cref{thm:main_result} seems to be a straightforward application of the universal approximation property of \acp{gmm}.
However, the \ac{gmm} sequence would then depend on \( \mby \) so that every new observation \( \mby \) would require a new sequence of \acp{gmm} to approximate the optimal estimator.
This is not desirable.
Instead, we only want one sequence \( (f_{\mbh}^{(K)})_{K=1}^\infty \) of \acp{gmm} which converges to \( f_{\mbh} \) and we show that this still implies the convergence of \( \hhat^{(K)} \) to \( \hhat \) for any given \( \mby \).

Since \( \mby \) is the sum of two random vectors, the \ac{pdf} \( f_{\mby} \) is the convolution of the \acp{pdf} \( f_{\mbh} \) and \( f_{\mbn} \).
Analogously, \( f_{\mby}^{(K)} \) is the convolution of \( f_{\mbh}^{(K)} \) and \( f_{\mbn} \).
A standard argument then shows that~\eqref{eq:convergence_fh} implies \( \lim_{K\to\infty} \| f_{\mby} - f_{\mby}^{(K)} \|_\infty = 0 \).
Next, we bound the quantity of interest:
\begin{align}
    \nonumber
    &\| \hhat - \hhat^{(K)} \|
    \leq
    \int \| \mbh \| \left| f_{\mbh\mid\mby}(\mbh\mid\mby) - f_{\mbh\mid\mby}^{(K)}(\mbh\mid\mby) \right| d \mbh
    \\&= \int \| \mbh \| | f_{\mbn}(\mby - \mbh) | \left| \frac{f_{\mbh}(\mbh)}{f_{\mby}(\mby)} - \frac{f_{\mbh}^{(K)}(\mbh)}{f_{\mby}^{(K)}(\mby)} \right| d \mbh
    \\&\leq \sup_{\mbh\in\C^N} \left| \frac{f_{\mbh}(\mbh)}{f_{\mby}(\mby)} - \frac{f_{\mbh}^{(K)}(\mbh)}{f_{\mby}^{(K)}(\mby)} \right| \int \| \mbh \| f_{\mbn}(\mby - \mbh) d \mbh.
    \label{eq:last_integral}
\end{align}
For the equality, we used~\eqref{eq:fh_given_y} and~\eqref{eq:fh_given_y_K}.
The integral in~\eqref{eq:last_integral} can be shown to be finite for any given \( \mby \) so that \Cref{thm:main_result} is proved as soon as the supremum in~\eqref{eq:last_integral} is shown to converge to zero for \( K \to \infty \).
This takes a bit more effort but is ultimately a consequence of the uniform convergence of \( f_{\mbh}^{(K)} \) to \( f_{\mbh} \) and of \( f_{\mby}^{(K)} \) to \( f_{\mby} \).
Details will be presented in a future work.

\section{NUMERICAL EXPERIMENTS}

Numerical experiments using two different channel models are presented after a review of related algorithms.

\subsection{Channel Models}\label{sec:channel_models}

In all simulations, the noise covariance matrix is \( \mbSigma = \sigma^2 \mbI \in \C^{N\times N} \).
Generated channels are normalized such that \( \expec[\| \mbh \|^2] = N \) holds which allows us to define a \ac{snr} as \( \mathrm{SNR} = 1 / \sigma^2 \).
The number of antennas is \( N = 128 \), and we generate \( 190 \cdot 10^3 \) training (e.g., for fitting a \ac{gmm} using an \ac{em} algorithm) and \( 10 \cdot 10^3 \) testing samples.
Unless stated otherwise, the \ac{gmm}-based estimator uses \( K = 128 \) components. 

\subsubsection{3GPP}\label{sec:3gpp}

We work with a spatial channel model~\cite{3gpp} where channels are modeled conditionally Gaussian: \( \mbh \mid \mbdelta \sim \calN(\mbzero, \mbC_{\mbdelta}) \).
The random vector \( \mbdelta \) collects the angles of arrivals and path gains of the main propagation clusters between a mobile terminal and the base station.
The base station employs a \ac{ula} such that the channel covariance matrix can be computed as
\( \mbC_{\mbdelta} = \int_{-\pi}^\pi g(\theta; \mbdelta) \mba(\theta) \mba(\theta)^\herm d \theta \).
Here, \( \mba(\theta) = [1, \exp(j\pi\sin(\theta), \dots, \exp(j\pi(N-1)\sin(\theta)]^\tp \) is the array steering vector for an angle of arrival \( \theta \) and \( g \) is a power density consisting of a sum of weighted Laplace densities whose standard deviations describe the angle spread of the propagation clusters~\cite{3gpp}.
For every channel sample, we generate random angles and path gains \( \mbdelta \) and then draw the sample as \( \mbh \sim \calN(\mbzero, \mbC_\mbdelta) \).

\subsubsection{QuaDRiGa}\label{sec:quadriga}

Version 2.4 of the QuaDRiGa channel simulator \cite{QuaDRiGa1, QuaDRiGa2} is used to generate channel samples for the numerical experiments.
We simulate an urban macrocell single carrier scenario at a frequency of 2.53 GHz.
The base station is equipped with a \ac{ula} with 128 ``3GPP-3D'' antennas and the mobile terminals employ an ``omni-directional'' antenna.
The base station is placed at a height of 25 meters and covers a sector of \( 120^\circ \).
The minimum and maximum distances between the mobile terminals and the base station are 35 meters and 500 meters, respectively.
In 80\% of the cases, the mobile terminals are located indoors at different floor levels, whereas the mobile terminals' height is 1.5 meters in the case of outdoor locations.

QuaDRiGa models channels as \( \mbh = \sum_{\ell=1}^{L} \mbg_{\ell} e^{-2\pi j f_c \tau_{\ell}} \),
where \( \ell \) is the path number, and the number of multi-path components $L$ depends on whether there is \ac{los}, \ac{nlos}, or \ac{o2i} propagation: \( L_\text{LOS} = 37 \), \( L_\text{NLOS} = 61 \) or \( L_\text{O2I} = 37 \).
The carrier frequency is denoted by \( f_c \) and the \( \ell \)-th path delay by \( \tau_{\ell} \).
The coefficients vector \( \mbg_{\ell} \) consists of one complex entry for each antenna pair, which comprises the attenuation of a path, the antenna radiation pattern weighting, and the polarization \cite{KuDaJaTh19}.
As described in the QuaDRiGa manual, the generated channels are post-processed to remove the path gain.

\begin{figure}[t]
	\centering
	\begin{tikzpicture}
		\begin{axis}
			[width=\plotwidth,
			height=\plotheightgpp,
			xtick=data,
			xmin=-15, 
			xmax=40,
			xlabel={SNR [dB]},
			ymode = log, 
			ymin= 1e-5,
			ymax=1,
			ylabel= {Normalized MSE}, 
			ylabel shift = 0.0cm,
			grid = both,
			legend columns = 2,
			legend entries={
				\legendAmp,
				\legendCnn,
				\legendGenielmmse,
				\legendGlobalcov,
				\legendGmm,
				\legendLs,
				\legendOmp,
			},
			legend style={at={(0.0,0.0)}, anchor=south west},
			]

			\addplot[amp]
			table[x=SNR, y=amp, col sep=comma]
			{icasspcsv/sim_55_results.csv};

			\addplot[cnn]
			table[x=SNR, y=cnn_fft2x_relu_non_hier_False, col sep=comma]
			{icasspcsv/cnn_1paths_128antennas_20lbatch_20lsize_500ebatch_2sigma_update.csv};

			\addplot[genielmmse]
			table[x=SNR, y=genie lmmse, col sep=comma]
			{icasspcsv/sim_55_results.csv};

			\addplot[globalcov]
			table[x=SNR, y=global cov, col sep=comma]
			{icasspcsv/sim_55_results.csv};

			\addplot[gmm]
			table[x=SNR, y=gmm full all, col sep=comma]
			{icasspcsv/sim_55_results.csv};
			
			\addplot[ls]
			table[x=SNR, y=ls, col sep=comma]
			{icasspcsv/sim_55_results.csv};
			
			\addplot[omp]
			table[x=SNR, y=Genie_OMP, col sep=comma]
			{icasspcsv/cnn_1paths_128antennas_20lbatch_20lsize_500ebatch_2sigma_update.csv};
			
		\end{axis}
	\end{tikzpicture}
	\caption{Channel estimation using the model from~\Cref{sec:3gpp} with one propagation cluster.}
	\label{fig:1path}
\end{figure}
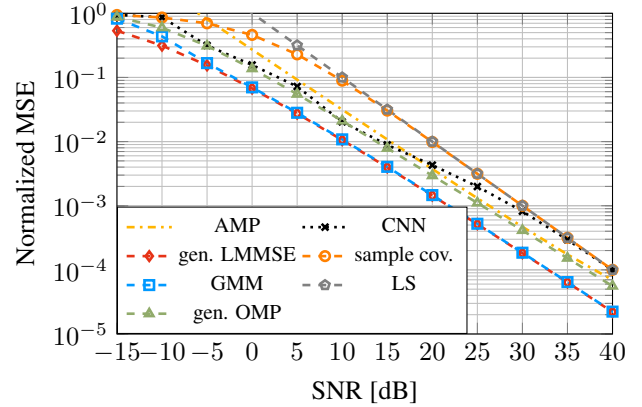

\begin{figure}[t]
	\centering
	\begin{tikzpicture}
		\begin{axis}
			[width=\plotwidth,
			height=\plotheightgpp,
			xtick=data,
			xmin=-15, 
			xmax=40,
			xlabel={SNR [dB]},
			ymode = log, 
			ymin= 5e-5,
			ymax=1,
			ylabel= {Normalized MSE}, 
			ylabel shift = 0.0cm,
			grid = both,
			legend columns = 2,
			legend entries={
				\legendAmp,
				\legendCnn,
				\legendGenielmmse,
				\legendGlobalcov,
				\legendGmm,
				\legendLs,
				\legendOmp,
			},
			legend style={at={(0.0,0.0)}, anchor=south west},
			]

			\addplot[amp]
			table[x=SNR, y=amp, col sep=comma]
			{icasspcsv/sim_56_results.csv};

			\addplot[cnn]
			table[x=SNR, y=cnn_fft2x_relu_non_hier_False, col sep=comma]
			{icasspcsv/cnn_3paths_128antennas_20lbatch_20lsize_500ebatch_2sigma.csv};

			\addplot[genielmmse]
			table[x=SNR, y=genie lmmse, col sep=comma]
			{icasspcsv/sim_56_results.csv};

			\addplot[globalcov]
			table[x=SNR, y=global cov, col sep=comma]
			{icasspcsv/sim_56_results.csv};

			\addplot[gmm]
			table[x=SNR, y=gmm full all, col sep=comma]
			{icasspcsv/sim_56_results.csv};
			
			\addplot[ls]
			table[x=SNR, y=ls, col sep=comma]
			{icasspcsv/sim_56_results.csv};
			
			\addplot[omp]
			table[x=SNR, y=Genie_OMP, col sep=comma]
			{icasspcsv/cnn_3paths_128antennas_20lbatch_20lsize_500ebatch_2sigma.csv};
			
		\end{axis}
	\end{tikzpicture}
	\caption{Channel estimation using the model from~\Cref{sec:3gpp} with three propagation clusters.}
	\label{fig:3paths}
	\vspace{-2mm}
\end{figure}
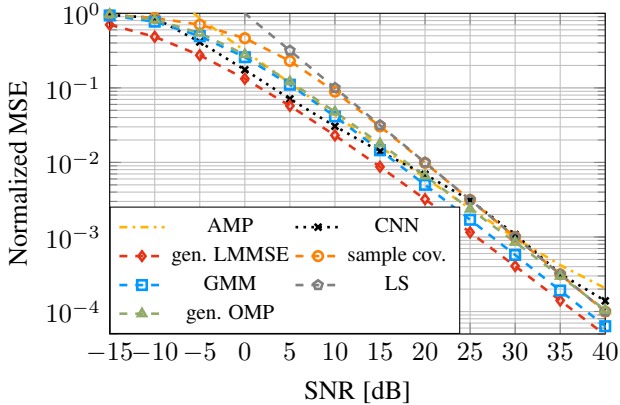

\subsection{Related Channel Estimators}

Two obvious baseline channel estimators are the \ac{ls} estimator which simply computes \( \hhat_{\text{LS}} = \mby \) in our case and the \ac{lmmse} formula based on the sample covariance matrix.
For the latter, we use all \( M = 190 \cdot 10^3 \) training data to estimate a sample covariance matrix \( \mbC = \frac{1}{M} \sum_{m=1}^{M} \mbh_m \mbh_m^\herm \) and then estimate channels as \( \hhat_{\text{sample cov.}} = \mbC (\mbC + \mbSigma)^{-1} \mby \).

In compressive sensing, the channel is assumed to be (approximately) sparse such that we have \( \mbh \approx \mbD \mbs \) for a sparse vector \( \mbs \in \C^L \).
The \textit{dictionary} \( \mbD \in \C^{N\times L} \) is typically an oversampled discrete Fourier transform matrix (e.g., \cite{AlLeHe15}).
Compressive sensing algorithms then recover an estimate \( \hat{\mbs} \) of \( \mbs \) assuming \( \mby = \mbD \mbs + \mbn \) and estimate the channel as \( \hhat = \mbD \hat{\mbs} \).
One algorithm is \ac{omp} \cite{DaMaAv97,PaReKr93,Tr04} which we employ with \( L = 4 N \).
\ac{omp} needs to know the sparsity order.
Since order estimation is a difficult problem, we avoid it via a genie-aided approach: \ac{omp} gets access to the true channel to choose the optimal sparsity.
This yields a performance bound for \ac{omp}.
A more elaborate compressive sensing algorithm is \ac{amp} \cite{DoMaMo10,MaAnYaBa13} which does not need to know the sparsity order.
We achieved the best results with \( L = 2 N \).

A \ac{cnn}-based channel estimator was introduced in~\cite{NeWiUt18}.
The authors exploit knowledge about the \ac{ula} geometry and about the 3GPP channel model in order to derive its architecture.
We use this \ac{cnn} as described in~\cite{NeWiUt18} for both channel models introduced in~\Cref{sec:channel_models}.
The \ac{cnn} is always trained on data that corresponds to the channel model on which it is tested afterwards.
The activation function is the rectified linear unit and we use the input transform based on the \( 2N \times 2N \) Fourier matrix, cf. \cite[Equation (43)]{NeWiUt18}.

\subsection{Numerical Results}
%

\Cref{fig:1path,fig:3paths} show the estimation performance in terms of normalized \ac{mse} when the channel model from \Cref{sec:3gpp} is used.
Since for every channel, a random covariance matrix \( \mbC_{\mbdelta} \) is generated and the channel is then drawn according to \( \mbh \sim \calN(\mbzero, \mbC_{\mbdelta}) \), we can use this channel covariance matrix to compute \ac{lmmse} channel estimates as \( \hhat_{\text{gen. LMMSE}} = \mbC_{\mbdelta} (\mbC_{\mbdelta} + \mbSigma)^{-1} \mby \) to provide a performance bound for all algorithms.
We call this estimator ``genie \ac{lmmse}'' because the true \( \mbC_{\mbdelta} \) is used which is not available in practice.

In \Cref{fig:1path}, we investigate channels with one propagation cluster.
It is interesting to see that the \ac{gmm}-based estimator performs almost as well as the genie \ac{lmmse} estimator.
In the mid-\ac{snr} range, the two compressive sensing algorithms are approximately equally good.
In \Cref{fig:3paths}, we have three propagation clusters.
A first observation is the strong performance of the \ac{cnn} estimator in the mid-\ac{snr} range.
Note that we can generally not expect any estimator to reach the genie \ac{lmmse} curve because it has more channel knowledge.
In the higher \ac{snr}-range, the \ac{gmm}-based estimator seems to be the only algorithm still outperforming \ac{ls} estimation.

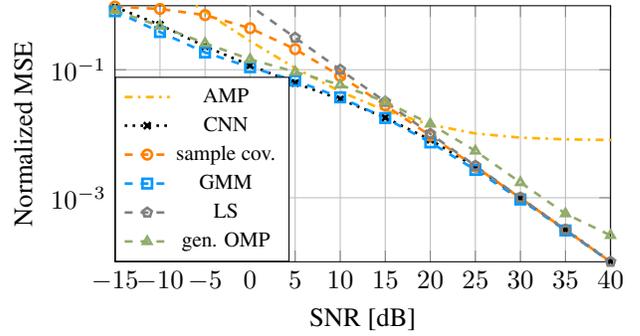
\begin{figure}[t]
	\centering
	\begin{tikzpicture}
		\begin{axis}
			[width=\plotwidth,
			height=\plotheightquadriga,
			xtick=data,
			xmin=-15, 
			xmax=40,
			xlabel={SNR [dB]},
			ymode = log, 
			ymin= 1e-4,
			ymax=1,
			ylabel= {Normalized MSE}, 
			ylabel shift = 0.0cm,
			grid = both,
			legend columns = 1,
			legend entries={
				\legendAmp,
				\legendCnn,
				\legendGlobalcov,
				\legendGmm,
				\legendLs,
				\legendOmp,
			},
			legend style={at={(0.0,0.0)}, anchor=south west},
			]

			\addplot[amp]
			table[x=SNR, y=amp, col sep=comma]
			{icasspcsv/sim_74_results.csv};

			\addplot[cnn]
			table[x=SNR, y=cnn_fft2x_relu_non_hier_False, col sep=comma]
			{icasspcsv/cnn_128antennas_quadriga.csv};

			\addplot[globalcov]
			table[x=SNR, y=global cov, col sep=comma]
			{icasspcsv/sim_74_results.csv};
			
			\addplot[gmm]
			table[x=SNR, y=gmm full all, col sep=comma]
			{icasspcsv/sim_74_results.csv};
			
			\addplot[ls]
			table[x=SNR, y=ls, col sep=comma]
			{icasspcsv/sim_74_results.csv};
			
			\addplot[omp]
			table[x=SNR, y=Genie_OMP, col sep=comma]
			{icasspcsv/cnn_128antennas_quadriga.csv};
			
		\end{axis}
	\end{tikzpicture}
	\caption{Channel estimation using the model from~\Cref{sec:quadriga}.}
	\label{fig:quadriga}
\end{figure}

In \Cref{fig:quadriga}, we concentrate on the QuaDRiGa channel model described in \Cref{sec:quadriga} where the channel covariance matrices and therefore the genie \ac{lmmse} curve are no longer available.
Here, the two compressive sensing algorithms behave not as similarly as they did in the previous experiments.
Additionally, their performance is not as convincing.
A reason might be that the channels now are not sparse enough.
The \ac{cnn} estimator shows again a good performance and overall the \ac{gmm}-based estimator can compete with it or is better.

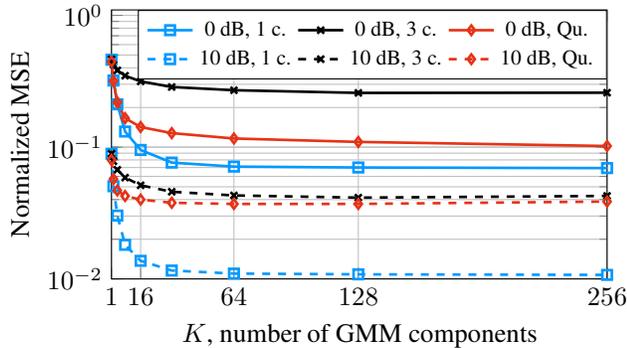
\begin{figure}[ht]
	\centering
	\begin{tikzpicture}
		\begin{axis}
			[width=\plotwidth,
			height=\plotheightcomponents,
			xtick=data,
			xmin=1, 
			xmax=256,
			xlabel={\( K \), number of \ac{gmm} components},
			xtick={1, 16, 64, 128, 256},
			ymode = log, 
			ymin= 1e-2,
			ymax=1.1,
			ylabel= {Normalized MSE}, 
			ylabel shift = 0.0cm,
			grid = both,
			legend columns = 3,
			legend entries={
                \footnotesize{0 dB, 1 c.},
                \footnotesize{0 dB, 3 c.},
                \footnotesize{0 dB, Qu.},
                \footnotesize{10 dB, 1 c.},
                \footnotesize{10 dB, 3 c.},
                \footnotesize{10 dB, Qu.},
			},
			legend style={at={(1.0,1.0)}, anchor=north east},
			]

			\addplot[onePlow]
			table[x=components, y=snr_0, col sep=comma]
			{icasspcsv/components_gpp_1path.csv};

			\addplot[threePlow]
			table[x=components, y=snr_0, col sep=comma]
			{icasspcsv/components_gpp_3paths.csv};

			\addplot[quaDlow]
			table[x=components, y=snr_0, col sep=comma]
			{icasspcsv/components_quadriga.csv};

			\addplot[onePhigh]
			table[x=components, y=snr_10, col sep=comma]
			{icasspcsv/components_gpp_1path.csv};

			\addplot[threePhigh]
			table[x=components, y=snr_10, col sep=comma]
			{icasspcsv/components_gpp_3paths.csv};

			\addplot[quaDhigh]
			table[x=components, y=snr_10, col sep=comma]
			{icasspcsv/components_quadriga.csv};

		\end{axis}
	\end{tikzpicture}
	\caption{Varying \( K \) for two different \acp{snr}. Channel models with one cluster (1 c.), three clusters (3 c.), QuaDRiGa (Qu.).}
	\label{fig:components}
	\vspace{-2mm}
\end{figure}

\Cref{fig:components} displays the last experiment where we analyze the effect of varying the number \( K \) of \ac{gmm} components.
The expected improvement of the \ac{mse} as \( K \) increases can be seen.
For \( K = 1 \), the \ac{gmm}-based estimator shows a performance almost identical to the sample covariance matrix-based estimator because the sample covariance matrix is exactly the maximum likelihood estimate in this case.
We emphasize that for larger \( K \), more training data is in general necessary to reach the \ac{gmm}'s performance limit because the \ac{gmm} has more parameters for larger \( K \).
This reflects a typical trade-off between performance and complexity, which may also depend on the amount of available training data.

\section{OUTLOOK}

We have presented preliminary results of the \ac{gmm}-based estimator in a simple channel estimation scenario that is representative of future applications for characterizing measurement campaigns or simulation tools.
In a future work, a rigorous proof of \Cref{thm:main_result} will be presented and we plan to analyze in particular the convergence behavior of the estimator.
Also, we intend to generalize the proof to multiple-input multiple-output channel estimation where a pilot matrix is present.


\bibliographystyle{IEEEbib}
\bibliography{IEEEabrv,references}

\end{document}

%% file: miko.tex
\usepackage{mathtools}	
\usepackage{amssymb}
\usepackage{amsthm}
\usepackage{siunitx}

\usepackage[utf8]{inputenc}

\usepackage{ifthen}

\usepackage{microtype}

\usepackage{bm}

\usepackage[draft,nomargin,marginclue,inline]{fixme}
\makeatletter
\renewcommand*\FXLayoutMarginClue[3]{%
  \marginpar[%
  \raggedleft\@fxuseface{margin}\textcolor{red}{\ignorespaces $ \Rightarrow $}]{%
    \raggedright\@fxuseface{margin}\textcolor{red}{\ignorespaces $ \Leftarrow $}}}
\makeatother	

\usepackage{tumcolor}

\usepackage{pgfplotstable}	

\pgfplotsset{
	discard if/.style 2 args={
        x filter/.append code={
            \edef\tempa{\thisrow{#1}}
            \edef\tempb{#2}
            \ifx\tempa\tempb
                
            \fi
        }
    },
    discard if not/.style 2 args={
        x filter/.append code={
            \edef\tempa{\thisrow{#1}}
            \edef\tempb{#2}
            \ifx\tempa\tempb
            \else
                
            \fi
        }
    }
}
\usepackage{cite}
\usepackage{hyperref}
\hypersetup{
    colorlinks = true,
    linkbordercolor = {white},
    citecolor = {black},
    linkcolor = {black},
    urlcolor = {black}
}
\makeatletter
\renewcommand*{\eqref}[1]{%
  \hyperref[{#1}]{\textup{\tagform@{\ref*{#1}}}}%
}
\makeatother

\usepackage[acronym,shortcuts]{glossaries}

\usepackage{cleveref}

\usepackage{algorithm}
\usepackage{algorithmic}

\usepackage[normalem]{ulem}

\pgfplotscreateplotcyclelist{miko}{%
{color=TUMBeamerRed, dash pattern=on 5pt off 1pt on 1pt off 1pt, line width=1.5pt},%
{color=TUMBeamerOrange, dash pattern=on 5pt off 2pt, line width=1.5pt,
mark=o},%
{color=black, dash pattern=on 5pt off 2pt on 3pt off 2pt, line width=1.5pt},%
{color=TUMDarkerBlue, line width=1.5pt},%
{color=black, dotted, line width=1.5pt},%
{color=TUMBeamerGreen, dotted, line width=1.5pt},%
{color=TUMBeamerYellow, dotted, line width=1.5pt},%
{color=TUMBeamerDarkRed, dotted, line width=1.5pt},%
{color=TUMBeamerLightGreen, dotted, line width=1.5pt},%
{color=TUMIvory, dotted, line width=1.5pt},%
{color=TUMLightBlue, dotted, line width=1.5pt},%
}

\DeclareMathOperator{\expec}{E}

\newcommand*{\C}{\mathbb{C}}

\newcommand{\herm}{{\operatorname{H}}}
\newcommand{\tp}{{\operatorname{T}}}

\newlength{\leftstackrelawd}
\newlength{\leftstackrelbwd}
\def\leftstackrel#1#2{\settowidth{\leftstackrelawd}%
	{${{}^{#1}}$}\settowidth{\leftstackrelbwd}{$#2$}%
	\addtolength{\leftstackrelawd}{-\leftstackrelbwd}%
	\leavevmode\ifthenelse{\lengthtest{\leftstackrelawd>0pt}}%
	{\kern-.5\leftstackrelawd}{}\mathrel{\mathop{#2}\limits^{#1}}}

\newcommand{\mbC}{\bm{C}}
\newcommand{\mbD}{\bm{D}}

\newcommand{\mbI}{\bm{I}}

\newcommand{\mba}{\bm{a}}

\newcommand{\mbg}{\bm{g}}
\newcommand{\mbh}{\bm{h}}

\newcommand{\mbn}{\bm{n}}

\newcommand{\mbs}{\bm{s}}

\newcommand{\mby}{\bm{y}}

\newcommand{\mbSigma}{{\bm{\Sigma}}}

\newcommand{\mbdelta}{{\bm{\delta}}}

\newcommand{\mbmu}{{\bm{\mu}}}

\newcommand{\mbzero}{{\bm{0}}}

\newcommand{\calH}{\mathcal{H}}

\newcommand{\calN}{\mathcal{N}}